\documentclass[10pt]{iopart}
\usepackage[english]{babel}
\usepackage[utf8]{inputenc}
\usepackage[plainpages=false,pdfpagelabels,unicode]{hyperref}
\usepackage[pdftex]{graphicx}
\usepackage{multirow}
\usepackage{xcolor}

\newcommand{\change}[1]{{\color{orange}#1}}
\newcommand{\changee}[1]{{\color{orange}#1}}

\begin{document}

\title{Accurate prediction of band gaps and optical properties of HfO$_2$}

\author{Pavel Ondračka$^{1,2}$, David Holec$^{3}$, David Nečas$^{1,2}$, and Lenka Zajíčková$^{1,2}$}

\address{$^1$ Department of Physical Electronics, Faculty of Science, Masaryk University, Kotlářská 2, 611 37 Brno, Czech Republic}
\address{$^2$ Plasma Technologies, CEITEC - Central European Institute of Technology, Masaryk University, Kotlářská 2, 611 37 Brno, Czech Republic}
\address{$^3$ Department of Physical Metallurgy and Materials Testing, Montanuniversität Leoben, Franz-Josef-Straße 18, Leoben A-8700, Austria}

\begin{abstract}
We report on optical properties of various polymorphs of hafnia predicted within the framework of density functional theory.
The full potential linearised augmented plane wave method was employed together with the Tran-Blaha modified Becke-Johnson potential (TB-mBJ) for exchange and local density approximation for correlation.
Unit cells of monoclinic, cubic, and tetragonal crystalline, and a simulated annealing-based model of amorphous hafnia were fully relaxed with respect to internal positions and lattice parameters.
\change{
Electronic structures and band gaps for monoclinic, cubic, tetragonal and amorphous hafnia were calculated using three different TB-mBJ parametrisation and the results were critically compared with available experimental and theoretical reports.
Conceptual differences between a straightforward comparison of experimental measurements to a calculated band gap on the one hand and to a whole electronic structure (density of electronic states) on the other hand, were pointed out, suggesting the latter should be used whenever possible.
Finally, dielectric functions were calculated at two levels, using the random phase approximation without local field effects and with a more accurate Bethe-Salpether equation (BSE) to account for excitonic effects.
We conclude that a satisfactory agreement with experimental data for HfO$_2$ was obtained only in the latter case.
}
\end{abstract}


\submitto{\JPD}

\ioptwocol

\section{Introduction}
\begin{figure}[b]
   \begin{center}
   \includegraphics[width=0.3\linewidth]{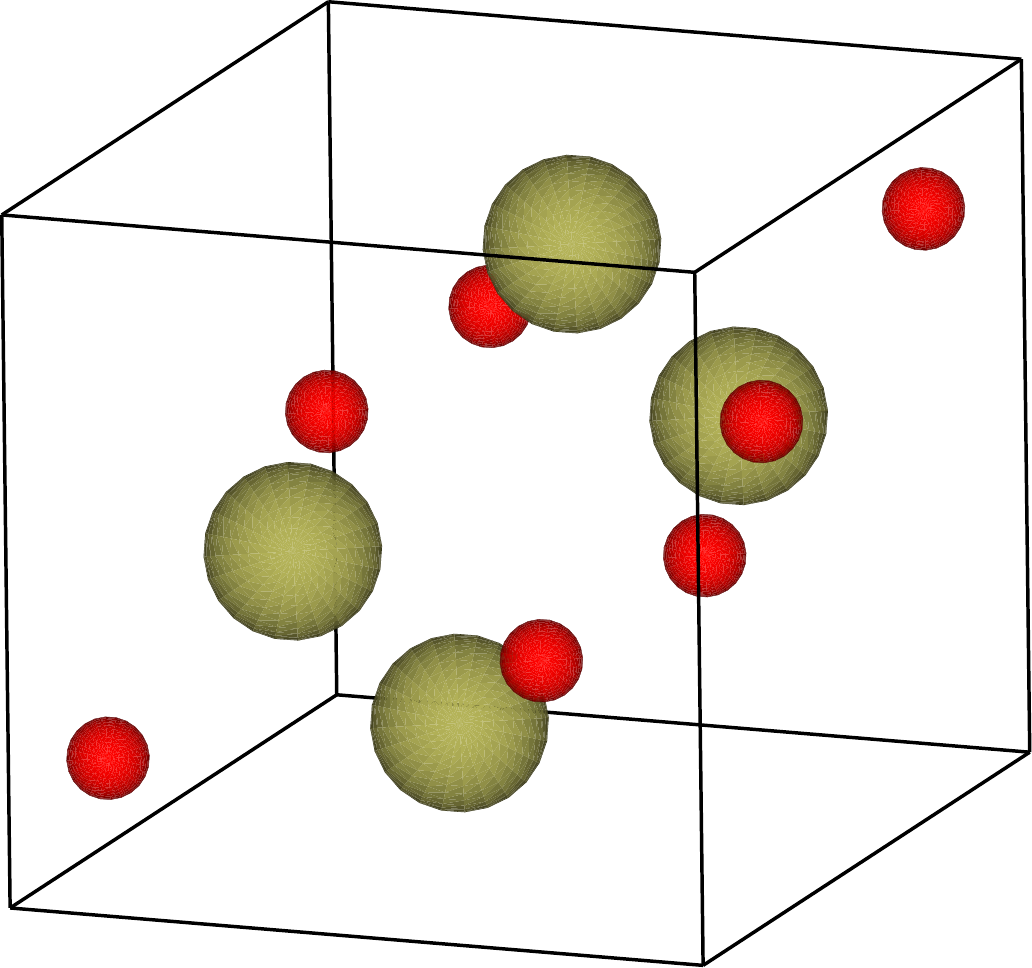}
   \includegraphics[width=0.3\linewidth]{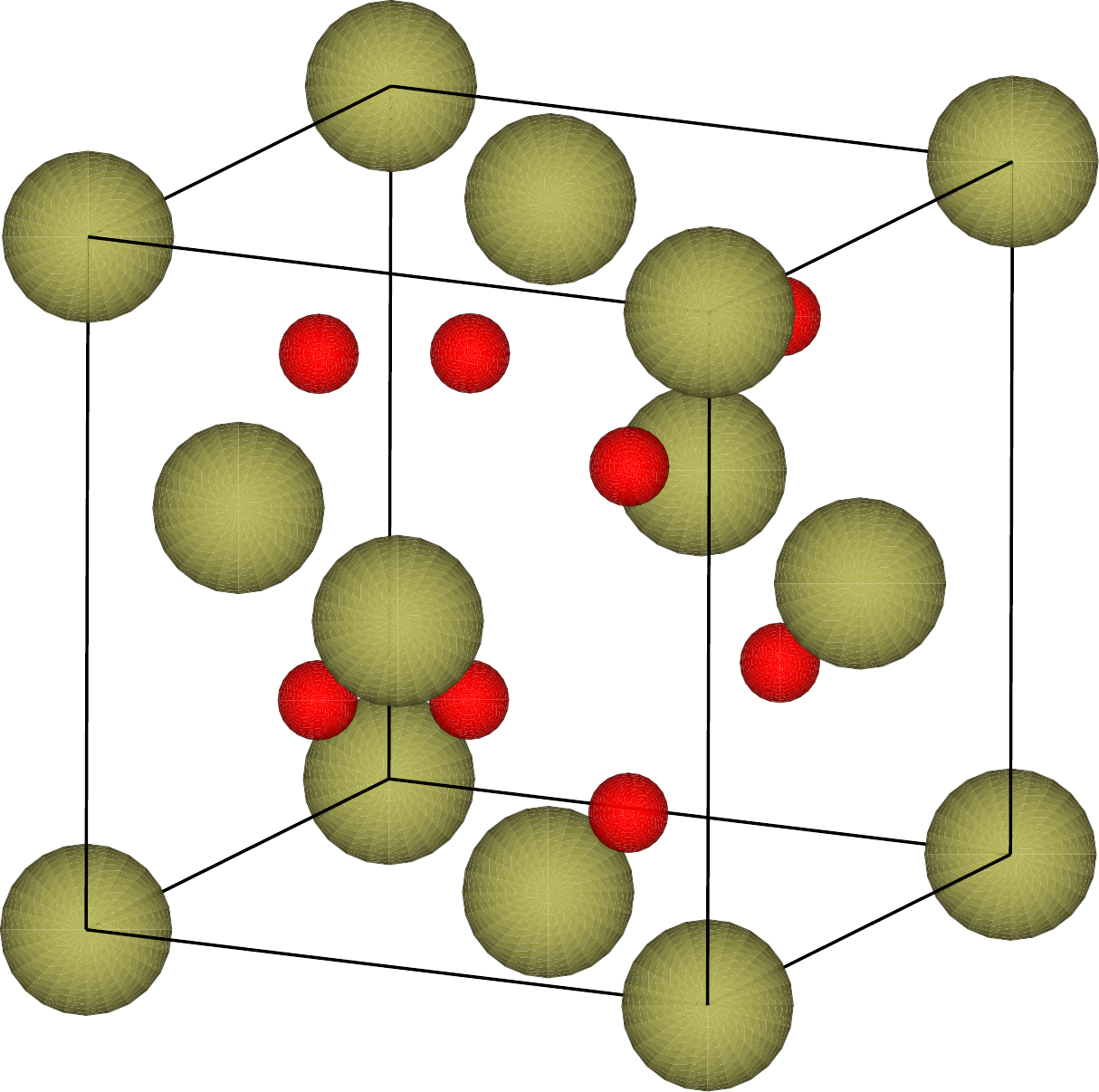}
   \includegraphics[width=0.3\linewidth]{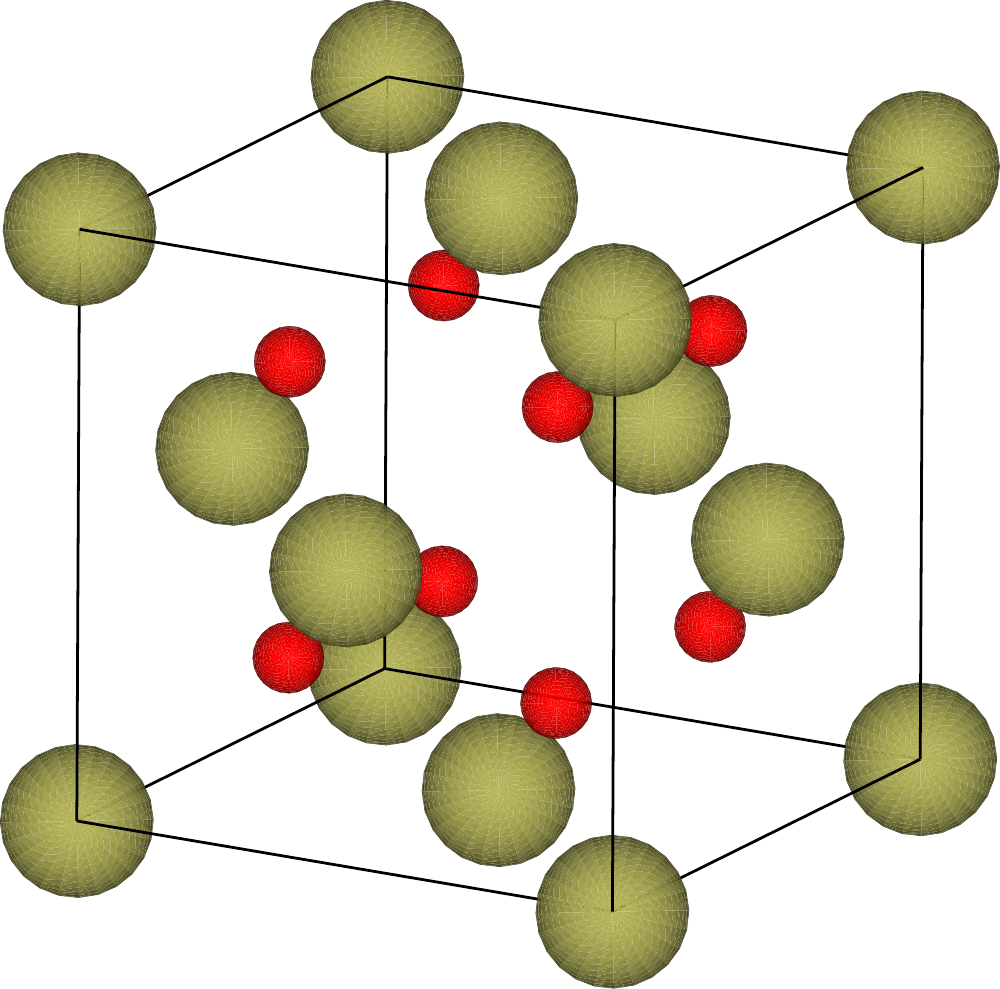}
   \caption{Crystal structure of a) monoclinic, b) tetragonal, and c) cubic hafnia, visualised by VESTA~\cite{Momma2011}.}
   \label{structs}
   \end{center}
\end{figure}

Hafnium dioxide (hafnia, HfO$_2$) is attracting a lot of attention as a perspective high-$k$ material for electronic~\cite{Houssa2006, Robertson2006} as well as optical applications such as antireflective coatings~\cite{Fadel1998, Khoshman2008}, or heat~\cite{Al-Kuhaili2004} and laser mirrors~\cite{Meng2012}.
According to the ambient pressure phase diagram, three crystalline polymorphs of HfO$_2$ are stable (Figure~\ref{structs}). The monoclinic structure ($P12_1/c1$, space group: \#14) is stable up to $\sim$1700\,$^\circ$C. The tetragonal phase ($P4_2/nmc$, space group: \#137) appears for temperatures between $\sim$1700\,$^\circ$C and $\sim$2220\,$^\circ$C,. At higher temperatures up to the melting point at $\sim2810\,^{\circ}$C, HfO$_2$ transforms to the cubic structure ($Fm\bar3m$, space group: \#225)~\cite{Villars2014-px}.

In addition to experimental works, there are numerous modelling, and in particular \textit{ab initio} studies reporting on structural, mechanical, electronic or optical properties of HfO$_2$~\cite{Caravaca2005, Broqvist2007, Ceresoli2006, Garcia2004, Kaneta2007, Liu2009, Scopel2008, Terki2008, Zhao2002, Gruning2010}.
Most of them employ conventional approximations for the exchange-correlation (xc) potential: local density approximation (LDA) or generalised gradient approximation (GGA).
A well known shortcoming of LDA and GGA is their underestimation of the band gap~\cite{Tran2009,Koller2011-hd,Koller2012}.

In order to predict band gaps in much better agreement with experiments, hybrid xc-functionals or other more complex methods (e.g., Green's functions-based GW approach) have been developed. However, these are significantly more computationally expensive in comparison with the simple LDA and GGA.
To overcome this difficulty, Tran and Blaha~\cite{Tran2009} have proposed a semi-local exchange potential (modified Becke-Johnson, TB-mBJ) that can provide highly accurate band gaps at the computational cost of LDA or GGA.

Using the original TB-mBJ parametrisation, Koller et al.~\cite{Koller2012} calculated the band gap of monoclinic HfO$_2$ to be 5.83\,eV, which is in good agreement with the experimental value of 5.68\,eV \cite{Balog1977}.
However, it is not clear whether the predictive power of TB-mBJ for band gaps extends also to other polymorphs of HfO$_2$.
Additionally, since HfO$_2$ is valued for its optical properties, it is also important to test whether the electronic structure calculated with TB-mBJ xc-potential and yielding an improved band gap will also guarantee a reliable prediction of the \change{electronic part of the} dielectric function.

Therefore, in this work we critically assess the band gaps and optical properties of monoclinic, tetragonal, cubic, and amorphous hafnia calculated with the TB-mBJ potential.

\section{Methodology}

The initial HfO$_2$ monoclinic, tetragonal and cubic cells were structurally optimized with respect to internal positions and lattice parameters.
This was done using the Vienna Ab initio Simulation Package~\cite{Kresse1996}, with projector augmented pseudopotentials~\cite{Kresse1999} and using both GGA parametrized by Perdew, Burke and Ernzerhof (GGA-PBE)~\cite{Perdew1996}, and LDA for the xc-potential.
LDA and GGA were used because there is no exchange energy functional associated with the TB-mBJ potential~\cite{Tran2009}, hence it is not suitable for structural optimizations based on the total energy minimisation.
The number of $k$-points reflected the size of the modelled cell by keeping the product (number of $k$-points)$\cdot$(number of atoms) constant and equal to approximately 2500.
The plane wave cut-off energy of 500\,eV used for crystalline polymorphs was reduced to 300\,eV for the amorphous models.
Consequently, a total energy accuracy of several meV/atom was achieved.

Two amorphous unit cells were prepared by the simulated annealing procedure~\cite{Drabold2009-ad}.
In both cells 96 atoms were randomly distributed inside a cubic simulation box with a side 10.1503\,\AA{} corresponding to the mass density of 10.695\,g/cm$^3$.
An \textit{ab initio} molecular dynamics run at 5000\,K for 3\,ps with a time step of 3\,fs provided a thermally equilibrated distribution of the atoms inside the cell.
In the next step, the cell temperature was decreased to 0\,K in 100 (fast) or 1000 (slow) steps, each corresponding to 3\,fs.
Finally, the resulting models were structurally relaxed with respect to atom positions and cell volume (i.e., mass density).

The electronic and optical properties were calculated for the structurally optimised models using Wien2k, a full potential all electron code~\cite{Blaha2001} employing the linearised augmented plane wave method.
Dense $k$-grids of 17$\times$17$\times$16 for monoclinic, 24$\times$24$\times$16 for tetragonal, 27$\times$27$\times$27 for cubic, and 4$\times$4$\times$4 for amorphous structures were used in order to obtain converged optical properties.
Atomic spheres radii were set to almost matching spheres corresponding to approximately (depending on the polymorph) 1.7\,\AA{} for oxygen and 1.9\,\AA{} for hafnium.
The $R_\mathrm{mt} \cdot K_\mathrm{max}$ matrix size parameter was set to 9, roughly equivalent to 410\,eV plane wave cut-off energy.

The TB-mBJ was used for the exchange part and LDA for the correlation part of the xc-potential.
\change{Three different parametrisation for the TB-mBJ were evaluated.
The original parametrisation by Tran and Blaha~\cite{Tran2009} (TB-mBJ original, TB-mBJ-orig) and the two improved parametrisations by Koller et al~\cite{Koller2012} one obtained using a larger testing set of solids (TB-mBJ improved, denoted as Present in the original manuscript, TB-mBJ-imp) and one parametrisation optimized for solids with band gap up to 7\,eV (TB-mBJ semiconductor, TB-mBJ-semi).
The electronic dielectric functions were calculated using the optic code~\cite{AmbroschDraxl2006}, a part of the Wien2k package, which utilizes Random Phase Approximation (RPA) neglecting the local field effects.
A Lorentz broadening of 0.03\,eV was applied to the dielectric function for a better comparison with the room-temperature experimental data.
To calculate the dielectric functions including excitonic effects, the Wien2kBSE module~\cite{Laskowski2006,Laskowski2009} was used.
\changee{No scissor operator was used to shift the dielectric function.}
Due to higher computational costs of the BSE calculation, smaller $k$-grids of 12$\times$12$\times$12, 7$\times$7$\times$5, and 4$\times$4$\times$4 for cubic, tetragonal and monoclinic phases, respectively, were adopted for the BSE Hamiltonian, \changee{while $k$-grids of 24$\times$24$\times$24, 14$\times$14$\times$10, and 8$\times$8$\times$8 were used for calculation of screening}.
Six conduction and fifteen valence bands per HfO$_2$ were considered in the BSE Hamiltonian.
\changee{A Gaussian broadening of 0.2\,eV was applied to the BSE spectra.}
}

\section{Results and discussion}

\subsection{Structural properties}

\begin{table*}
\caption{\label{structure}Overview of calculated structural properties of monoclinic (m, m2), tetragonal (t), cubic (c) and amorphous (a) phases. $a$, $b$, $c$ denote unit cell lattice parameters, $\beta$ is the monoclinic angle, $B$ is bulk modulus, $E_\mathrm{f}$ is the energy of formation, $\Delta E_\mathrm{f}$ is the difference between $E_\mathrm{f}$ for a given phase and the monoclinic (m) phase (the lowest energy configuration), and $\rho$ is the mass density.}
\lineup
\begin{indented}
\item[]\begin{tabular}{llllllllll}

\br
 & & $a$\,[\AA] & $b$\,[\AA] & $c$\,[\AA] & $\beta\,[^{\circ}]$ & $\rho\,[\mathrm{g/cm^3}]$ & $B$\,[GPa] & $E_\mathrm{f}$\,[eV/atom] & $\Delta E$\,[eV/atom]\\
\mr

m & GGA & 5.143 & 5.190 & 5.330 & \099.6 & \09.952 & 193 & -3.841 & 0.000
\\
  & LDA & 5.035 & 5.123 & 5.198 & \099.6 & 10.577 & 194 & -4.318 & 0.000
\\
  & exp~\cite{Adam1959} & 5.1156 & 5.1722 & 5.2948 & \099.23 & 10.111 & & & \\
\mr

m2 & GGA & 5.609 & 5.707 & 5.908 & 115.9 & \08.242 & 180 & -3.832 &
0.009\\ 
   & LDA & 5.534 & 5.606 & 5.775 & 115.8 & \08.646 & 206 & -4.247 &
0.071\\ 
\mr

t & GGA & 3.593 & & 5.232 & \090.0 & 10.349 & 170 & -3.786 & 0.055\\ 
  & LDA & 3.526 & & 5.073 & \090.0 & 11.086 & 220 & -4.280 & 0.038\\ 
  & exp~\cite{Curtis1954} & 3.634 & & 5.25 & & 10.083 & & & \\
\mr

c & GGA & 5.075 & & & \090.0 & 10.682 & 250 & -3.752 & 0.088 \\
  & LDA & 4.982 & & & \090.0 & 11.305 & 289 & -4.261 & 0.057 \\
  & exp~\cite{Senft1983} & 5.110 & & & & 10.478 & & & \\
\mr


a & GGA, slow cooling & & & & & \09.570 & 124 & -3.676 & 0.165\\
a & GGA, fast cooling & & & & & \09.857 & 140 & -3.668 & 0.173\\
\br

\end{tabular}
\end{indented}
\end{table*}

\begin{figure}[b]
\begin{center}
	\includegraphics[width=\linewidth]{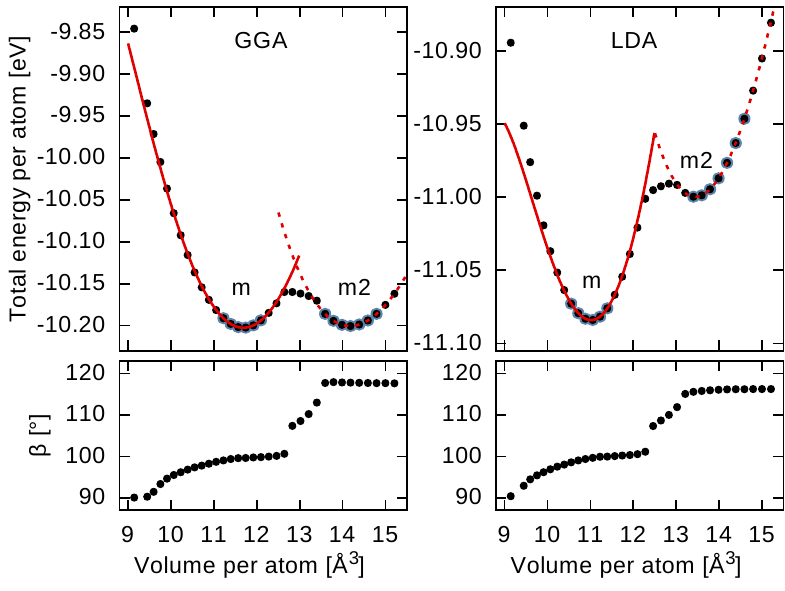}
	\caption{Total energy vs. volume data for the monoclinic polymorph calculated with GGA-PBE and LDA. Solid and dashed lines represent fits of the Birch-Murnaghan equation of state around centred at minima m and m2, respectively (only the highlighted points were fitted). The bottom panels show corresponding  monoclinic angle $\beta$ as a function of the specific volume.}
   \label{EV}
\end{center}
\end{figure}

The four structures considered in the present work (monoclinic, cubic, tetragonal, and amorphous) were fully optimised with respect to the unit cell shape and atom positions at several fixed volumes.
In doing so, we used both GGA and LDA exchange-correlation potentials that typically overestimate and underestimate, respectively, lattice parameters with respect to experimental values.
The thus obtained energy vs. volume data were fitted with Birch-Murnaghan equation of state~\cite{Birch1947}.
The resulting lattice parameters are listed in Table~\ref{structure} together with the selected experimental data from literature for comparison.
The energy of formation was calculated as
\begin{equation}
  E_\mathrm{f}^i=E_0(\mathrm{HfO}_2^i) - \frac13 (E_0(\mathrm{Hf}) + 2E_0(\mathrm{O_2}))\ ,
\end{equation}
where $E_0(\mathrm{HfO}_2^i)$, $E_0(\mathrm{Hf})$, and $E_0(\mathrm{O_2})$ are the \textit{ab initio} total energies (per atom) of HfO$_2$ in the structure $i$ (monoclinic, cubic, tetragonal, or amorphous), Hf in the hcp structure, and an oxygen molecule, respectively.

The most stable structure is, in agreement with the equilibrium phase diagram~\cite{Villars2014-px}, the monoclinic structure, followed by tetragonal (experimentally high temperature) and cubic structures.
The least stable configuration yielding the highest energy of formation is the amorphous structure.
The optimised lattice parameters are in good agreement with the previously published data, the GGA values being generally closer to the experimental values, hence only the GGA optimised structures were subsequently used in the TB-mBJ calculations.
The optimisation of the amorphous structure yielded estimates for the equilibrium mass density: $\rho_{\mathrm{slow}}=9.570\,\mathrm{g/cm^3}$, $\rho_{\mathrm{fast}}=9.857\,\mathrm{g/cm^3}$.
These values are slightly smaller than the optimised density of monoclinic structure m and the tetragonal and cubic structures are also predicted to be more dense (Table~\ref{structure}).

An interesting phenomenon is predicted for the monoclinic variant.
As shown in Figure~\ref{EV}, the energy vs. volume data exhibit a second minimum (quite a shallow one in the LDA case) for volumes larger than the equilibrium (structural parameters are denoted as m2 in Table~\ref{structure}).
The most distinct feature is a step-like change of the monoclinic angle $\beta$ from $99.6\,^\circ$ (GGA/LDA) to $115.9\,^\circ$ (GGA) and $115.8\,^\circ$ (LDA) (see bottom panels in Figure~\ref{EV}) \change{as well as a significantly lower mass density of the m2 phase as compared with all other HfO$_2$ polymorphs.
These two monoclinic structures also differ in the internal arrangements of their atoms.
The lower-energy state labelled m (the true ground state of HfO$_2$) is composed of two types of O environments: half of the O atoms are 3-coordinated (bonded to 3 Hf atoms), while the other half is 4-coordinated.
Consequently, all Hf atoms are 7-coordinated.
Such arrangement is in accordance with known structure from literature \cite{Zhang2013-sc}.
On the contrary, all O atom in the m2 structure have identical local environments: they are all 3-coordinated, leading to 6-coordination of all Hf atoms.
This lower coordination corresponds well with the lower density of the m2 phase.
}

A similar structural transition has been recently reported for the monoclinic phase in NiTi shape memory alloys \cite{Holec2011-tg}.
The structural complexity of the NiTi monoclinic phase resulted in a hysteresis of the forwards and backwards pressure-induced phase transformation.
\change{Therefore, we have calculated transformation pressures for HfO$_2$, defined as pressures at which the minimum of enthalpy is reached simultaneously by (at least) two different phases \cite{Holec2011-tg}.
The GGA data yield that the m$\to$m2 transition happens at $-0.025\,\mathrm{GPa}$ (i.e., under a tensile loading), and further m2$\to$t transformation is predicted at $-0.131\,\mathrm{GPa}$.
In contrast, the LDA data suggest only one transition, m$\to$t at $-0.059\,\mathrm{GPa}$.
The appearance of the m2 phase in the GGA evaluation is a direct consequence of its low energy of formation, comparable with that of the m phase.
It is worth mentioning that almost identical results were obtained also when the PW91 parametrisation of the GGA xc potential \cite{Wang1991-ca} was used instead of the PBE.

Despite the low $E_{\rm f}$ values for the m2 phase, it was not considered for the subsequent (computationally demanding) optical calculations.
The reasons for this decision are following: (i) it was predicted stable only by GGA and only at tensile isotropic stresses, (ii) it has significantly lower mass density than the experimentally observed HfO$_2$ materials, and (iii) it does not appear in the Hf--O phase diagram.
}

\subsection{Band gaps}

\begin{table*}
\caption{\label{gaps}Overview of predicted band gaps (in eV) compared with experimental data and other \textit{ab initio} calculations. Band gaps of amorphous structures were obtained using a Tauc plot approach, while the data in parentheses are electronic gaps as appearing in, e.g., total density of states (for details see text).}
\lineup
\begin{indented}
\item[]\begin{tabular}{l|llll|lll}
\br
HfO$_2$ & \multicolumn{4}{c|}{present study} & \multicolumn{3}{c}{literature reports}\\
phase & \multicolumn{3}{c|}{TB-mBJ} & PBE & hybrid functionals & GW$_0$ & experiment\\
 & orig. & imp. & semi. & & & & \\
\mr
m &	5.76 & 6.01 & 6.54 & 4.08 & PBE0: 6.75~\cite{Komsa2010}, HSE06: 5.98~\cite{Komsa2010} & 5.9~\cite{Gruning2010} & 5.68~\cite{Balog1977} \\
c &	5.88 & 6.17 & 6.74 & 3.77 & SX: 5.6~\cite{Clark2010}, HSE06: 5.38~\cite{Yang2014} & 5.5~\cite{Gruning2010} & 5.8~\cite{Lim2002}\\
t &	6.54 & 6.81 & 7.35 & 4.79 & & 6.0~\cite{Gruning2010} & \\
a$_{\rm slow}$ & 5.50 (5.52) & 5.75 (5.76) & 6.23 (6.25) & & \multirow{2}{*}{PBE0: 5.3~\cite{Broqvist2007}, 5.94~\cite{Chen2011}} &  & 5.49--5.72~\cite{Takeuchi2004},\\
a$_{\rm fast}$ & 5.23 (4.77) & 5.46 (4.98) & 5.98 (5.43) & & & & 5.62~\cite{Nguyen2005}, 5.7~\cite{Perevalov2007} \\
\br

\end{tabular}
\end{indented}
\end{table*}

The electronic structure of different HfO$_2$ polymorphs was calculated using the Wien2k code and \change{TB-mBJ parametrisations.}
The obtained band gaps are summarised in Table~\ref{gaps} together with experimental data and other \textit{ab initio} predictions.
\change{Values of 5.76\,eV, 6.01\,eV and 6.54\,eV were obtained for the band gap of m-HfO$_2$ with original, improved and semiconductor TB-mBJ parametrisations, respectively.
The TB-mBJ-orig value is in excellent agreement with the experimental value of 5.68\,eV~\cite{Balog1977} and a huge improvement over the PBE value of 4.08\,eV. The TB-mBJ-imp value is comparable with predictions employing the HSE06 hybrid xc-potential (5.98\,eV)~\cite{Komsa2010} or the GW$_0$ method (5.9\,eV)~\cite{Gruning2010}, while the TB-mBJ-semi value is more in line with 6.75\,eV predicted by the PBE0 hybrid functional~\cite{Komsa2010}.
Note that there is a slight ($\sim$0.1\,eV) disagreement between the TB-mBJ band gap values as reported by Koller \textit{et al.}~\cite{Koller2012} and our results.
We suggest that this is due to small differences in the unit cells.
If experimental structural values of Ruh and Corfield~\cite{Ruh1970} are used, the resulting TB-mBJ-orig band gap is 5.88\,eV.
The observed difference hence falls within the uncertainty caused by structural parameters.
It is unfortunately unclear if Koller \textit{et al.} used experimental lattice parameters or some optimisation procedure was applied similarly to this work.
}

\change{
The band gap of c-HfO$_2$ was estimated as 5.88\,eV with TB-mBJ-orig, which is a slightly higher value in comparison with hybrid functionals (SX: 5.6\,eV~\cite{Clark2010}, HSE06: 5.38\,eV~\cite{Yang2014}) as well as the GW$_0$ method (5.5\,eV~\cite{Gruning2010}).
Therefore, the TB-mBJ-imp 6.17\,eV and TB-mBJ-semi 6.74\,eV values, being even higher than that of TB-mBJ-orig, appear overestimated.
Comparison with an experiment is, however, not straightforward since cubic hafnia is not stable at room temperature, and hence it is usually stabilized by yttrium.
The band gap of 5.8\,eV~\cite{Lim2002} was reported for (Y$_2$O$_3$)$_{0.15}$(HfO$_2$)$_{0.85}$, which compares favourably with TB-mBJ-orig predictions.
It is worth noting that while in the case of PBE calculations the c-HfO$_2$ band gap is 0.3\,eV smaller than for m-HfO$_2$, the TB-mBJ potential predicts it to be 0.1--0.2\,eV larger.
It thus follows that the corrections induced by the TB-mBJ xc-potential depend both on the chemistry and structural properties.

The largest band gap of all hafnia polymorphs is predicted for the tetragonal structure, with values of 6.54, 6.81 and 7.35\,eV for orig, imp, and semi parametrisation, respectively.
This result is in agreement with the GW$_0$-based calculations, where the band gap of 6.0\,eV is also larger than those of monoclinic and cubic structures, albeit the differences are smaller.
Finally, all TB-mBJ calculated band gaps are significantly increased and improved towards experimental values in comparison with the GGA-PBE values of 4.08, 3.77, and 4.79\,eV for monoclinic, cubic and tetragonal HfO$_2$, respectively.
}

\begin{figure}
\begin{center}
	\includegraphics[width=\linewidth]{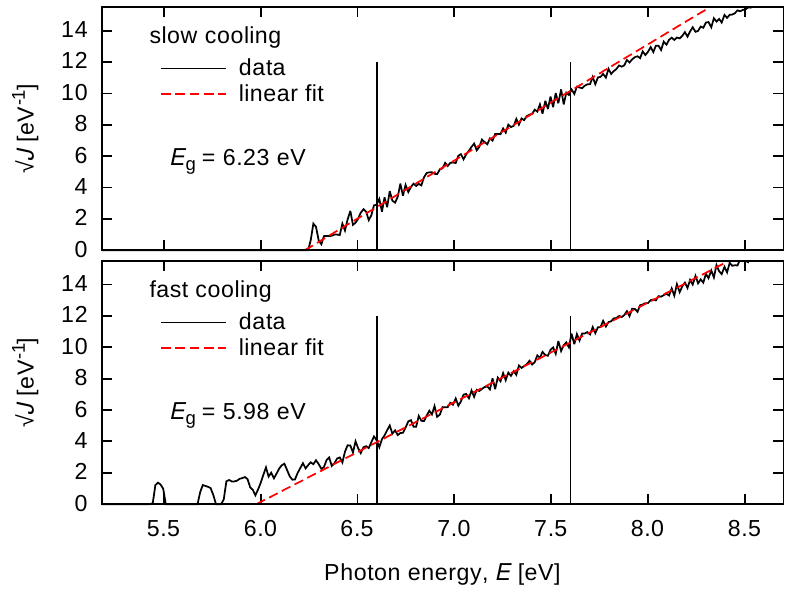}
	\caption{Tauc-like plot of $\sqrt{J(E)}$ used to estimate optical band gaps of the two structural models for amorphous structures for TB-mBJ-semi. The vertical solid lines denote the data range used for the fitting procedure.}
   \label{JDOS}
\end{center}
\end{figure}

\change{
While the above discussed band gaps energies of crystalline structures were determined simply as the differences between the energies of the highest occupied and lowest unoccupied states,
such simple calculation is not suitable for amorphous cells due to the possible presence of defect states appearing near the band gap edges.
Therefore, a different method was used for the amorphous cells,
with the goal of obtaining the fundamental band gaps and filtering out the defect states that depend strongly on local relaxations.
A similar approach is often applied to experimental optical measurements, in which the band gap upper energy is traditionally obtained by extrapolating the linear part of $\sqrt{\alpha E n}$ near the absorption onset (i.e., so-called Tauc plot), where $\alpha$ is the absorption coefficient, $E$ is the photon energy, and $n$ is the refractive index~\cite{Stenzel2005}.
The same approach has been already applied to simulated amorphous cells of TiO$_2$~\cite{Landmann2012}.
The optical band gap of amorphous cells was estimated by an equivalent procedure, fitting the linear part of $\sqrt{J(E)}$, where $J(E)$ is the joint density of occupied and unoccupied states as calculated by Wien2k.
This procedure filters out defects states around the edges of valence and conduction bands (see Figure~\ref{JDOS}).
Consequently, in the following we distinguish between Tauc-like and electronic (highest occupied--lowest unoccupied orbital) band gaps (cf. two values for the amorphous structures in Table~\ref{gaps}).
It can be concluded that in the case of slow cooling, the role of defect states on the band gap estimation is negligible, in contrast to the fast cooled amorphous cell.
We attribute it to the fact that more structural features (resulting in the defect electronic levels) corresponding to the high-temperature amorphous mixture were ``frozen in'' during the fast cooling process.
The predicted optical TB-mBJ-orig band gap of 5.50\,eV and TB-mBJ-imp band gap of 5.75\,eV for amorphous hafnia are in excellent agreement with the experimental values ranging from 5.49\,eV to 5.72\,eV~\cite{Takeuchi2004, Nguyen2005, Perevalov2007}.
The TB-mBJ-semi value of 6.25\,eV seems again slightly overestimated.
}

\change{
The band gap corrections introduced by the different employed TB-mBJ parametrisations were similar for all the calculated polymorphs. The band gap difference between TB-mBJ-orig and TB-mBJ-imp was $\sim$0.25\,eV and between TB-mBJ-orig and TB-mBJ-semi was $\sim$0.8\,eV regardless of the structure.
}

\begin{figure}
\begin{center}
	\includegraphics[width=\linewidth]{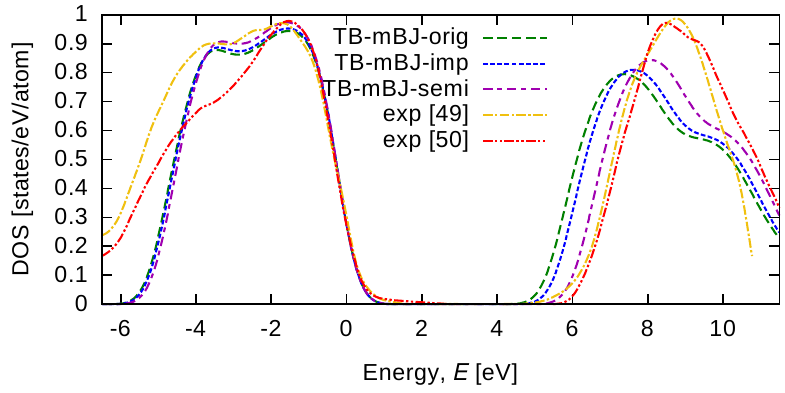}
	\caption{Total densities of states for a-HfO$_2,\mathrm{slow}$ compared with reported experimental data~\cite{Bersch2008,Sayan2004}. Experimental data were aligned by valance band edge and normalized to maximum valence band intensity. Gaussian broadening was applied to calculated results in order to match the experimental slope of the valence band edge. Zero energy corresponds to the top of the predicted valence band before broadening.}
   \label{am-DOS}
\end{center}
\end{figure}

\change{The simple comparison of band gaps performed above suggests that the original parametrisation TB-mBJ-orig is the best for HfO$_2$. 
Nonetheless, we note than such conclusion is misleading.
Band gaps obtained from optical measurements may be shifted down due to excitonic effects, and even in techniques that probe quasiparticle band gap such as photoelectron spectroscopy (PES) and inverse photoelectron spectroscopy (IPS), it may be difficult to distinguish the influence of imperfections in the sample, surface adsorbants, and substrate emission~\cite{Bersch2008}.
Moreover, the final value of the experimental band gap is usually highly dependent on a specific interpretation, especially the chosen region for fitting of the gap~\cite{Bersch2008}.
Therefore, we propose that a more rigorous testing of theoretical predictions against experiment on the level of the density of electronic states (DOS) should be performed instead.
The comparison of calculated and reported experimental DOS of a-HfO$_2$ measured by PES+IPS is shown in Figure~\ref{am-DOS}.
Obviously, the entire conduction band calculated with the TB-mBJ-orig and TB-mBJ-imp 
is shifted to lower energies with respect to the experimental DOS, while the semiconductor parametrisation yields the best prediction, with difference between calculated and experimental edge of the conduction band smaller than 0.5\,eV.
}

\subsection{Optical properties}

\begin{figure}
\begin{center}
	\includegraphics[width=\linewidth]{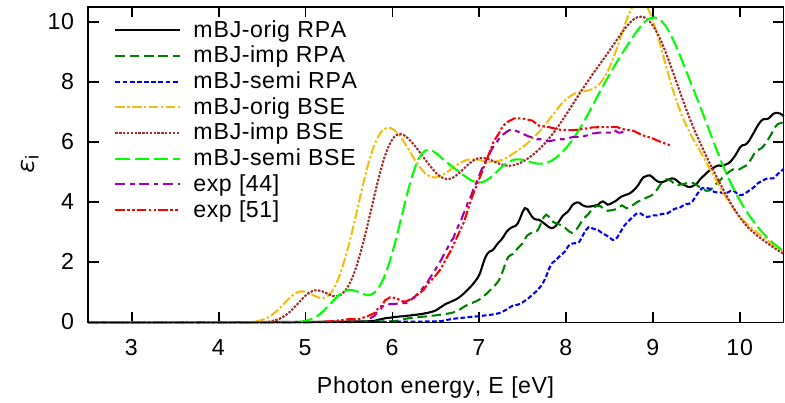}
	\caption{Comparison of calculated directionally averaged and experimental imaginary part of the dielectric function of monoclinic hafnia.}
   \label{m-eps-compare}
\end{center}
\end{figure}

\begin{figure*}[tbp]
\begin{center}
	\includegraphics[width=\linewidth]{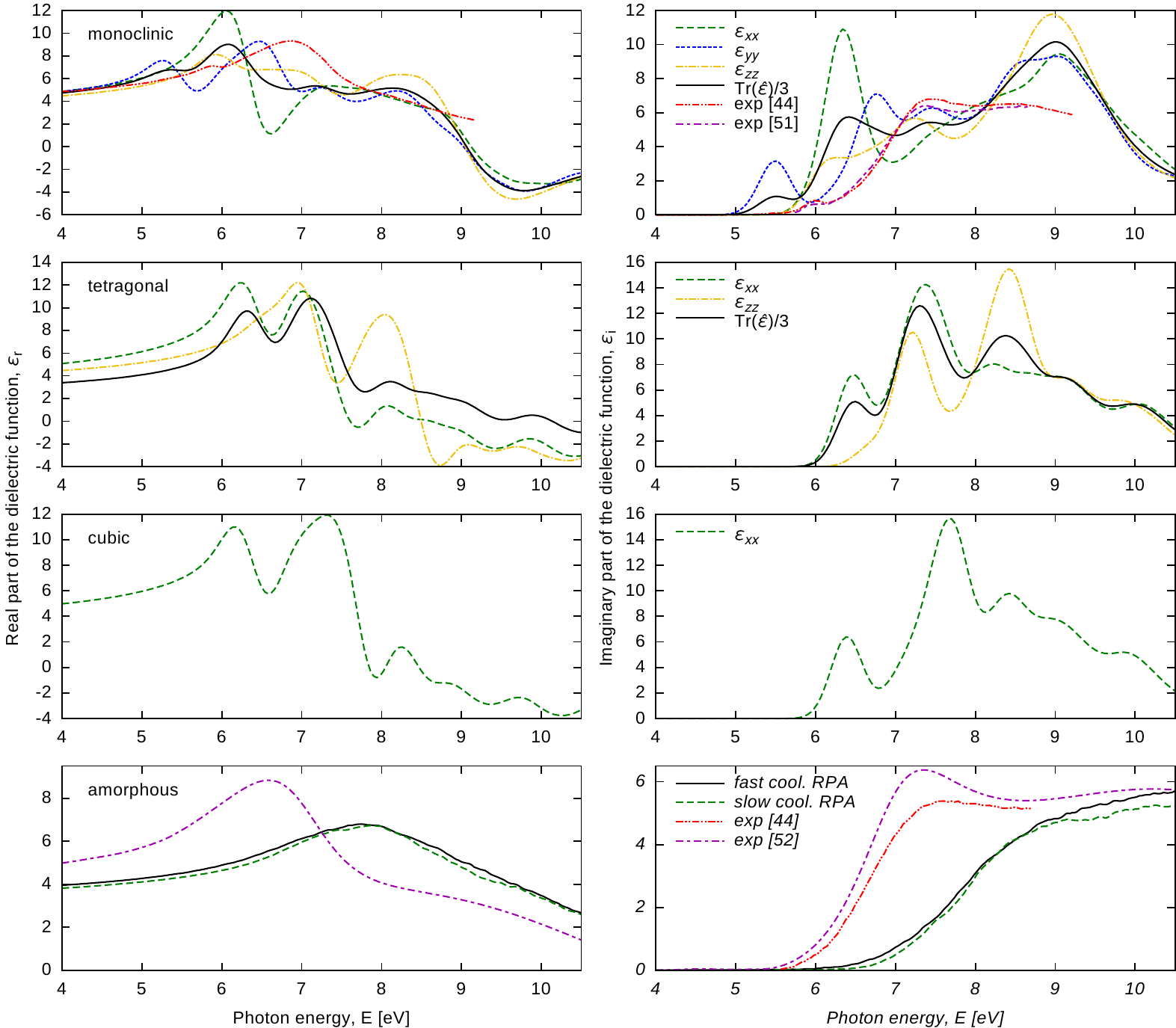}
	\caption{Calculated real (left) and imaginary (right) parts of the dielectric function of monoclinic, tetragonal, cubic and amorphous hafnia. Experimental data are added for comparison where available.}
   \label{eps-all}
\end{center}
\end{figure*}

\change{
The calculated optical properties are represented by the complex dielectric tensor, $\hat{\varepsilon}$.
The respective crystal symmetry implies that there is one component of the dielectric tensor $\hat{\varepsilon}$ for the cubic structure, two for the tetragonal, and four for the monoclinic hafnia.
Amorphous hafnia was treated as a crystal with only a primitive symmetry (i.e., identity) and three diagonal components were calculated.
However, they were found to be almost identical, and thus confirming the cell amorphousness.
Therefore, only the directionally averaged value calculated as $\Tr(\hat{\varepsilon})/3$ is shown.
The dielectric tensor is oriented so that the $xx$ component is parallel to the $a$ lattice vector, $yy$ lies in the plane defined by the $a$ and $b$ lattice vectors and it is orthogonal to the $xx$ component, and the $zz$ component is perpendicular to the $xx$ and $yy$ components.

The calculated imaginary part of the dielectric function ($\varepsilon_\mathrm{i}$) for different TB-mBJ parametrisations at the RPA and BSE levels is compared to experimental data for the monoclinic HfO$_2$ in Figure~\ref{m-eps-compare}.
Only the directionally averaged dielectric function, calculated as $\Tr(\hat{\varepsilon})/3$, is shown for a better comparison with experimental polycrystalline data.
It becomes immediately clear that excitonic effects play a significant role in the optical absorption. 
The imaginary part of dielectric function is significantly underestimated by RPA calculations and it lacks main features observed in the experimental spectra. 
No significant differences in the spectral features were observed between the different TB-mBJ parametrisations, and since the best overall agreement is obtained when using the TB-mBJ-semi parametrisation in the combination with BSE, we will further focus only on these data.

Figure~\ref{eps-all} summarises the real ($\hat{\varepsilon}_\mathrm{r}$) and imaginary ($\hat{\varepsilon}_\mathrm{i}$) parts of the dielectric tensor evaluated for the four discussed polymorphs of HfO$_2$. The $\hat{\varepsilon}$ of the three crystalline polymorphs was calculated using the TB-mBJ-semi parametrisation in the combination with BSE but these calculations were not computationally feasible for a-HfO$_2$ and, therefore, the results for amorphous HfO$_2$ correspond to TB-mBJ-semi together with RPA. 

Available experimental data from literature were added to Figure~\ref{eps-all} for a comparison.
Although the qualitative agreement is acceptable, there are still several differences.
The calculated imaginary parts of $\hat\varepsilon$ for m-HfO$_2$ are shifted to slightly lower energies with respect to the experimental datasets by Edwards~\cite{Edwards2003} and Nguyen \textit{et al.}~\cite{Nguyen2005}.
Importantly, we were able to reproduce the small peak at around 6\,eV appearing in experimental $\varepsilon_i$ of m-HfO$_2$, albeit shifted to lower enrgies by 0.5\,eV.
Previously, there was some uncertainty about this peak because Nguyen \textit{et al.}~\cite{Nguyen2005} attributed it to the absorption at defect states originating from oxygen vacancies, while Edwards~\cite{Edwards2003} suggested a possible excitonic origin.
Our results prove the second hypothesis, i.e., this peak is caused by a below-the-gap exciton.
\changee{The second peak (present around 7.3\,eV in experimental spectra) is shifted to lower energies in our prediction but its shape and intensity match the experiment almost perfectly.
An obvious disagreement between the calculations and the experimental data is the presence of the peak at 9\,eV in the predicted $\varepsilon_\mathrm{i}$.
This peak did not show up in any of the m-HfO$_2$ experimental spectra. Nevertheless, we suggest that it is caused by the limited spectral range.
The only available experimental data for a-HfO$_2$ in the range above 9\,eV, which has been published recently by Franta \textit{et al.}~\cite{Franta2015}, exhibit a broad peak centered at around 10.3\,eV.
Unfortunately, we cannot unambiguously conclude whether this peak is indeed a higher energy feature shifted to lower energies in our calculations, or if it results from TB-mBJ insufficiency to describe the band structure properly.
There are large differences between individual $\hat{\epsilon}$ components of m-HfO$_2$, especially near the absorption onset.
Most notably, the first excitonic peak is present only in the $yy$ component, while the second is most pronounced in the $xx$ component.

Similarly, the dielectric function of t-HfO$_2$ also exhibits a large difference between the two independent dielectric tensor components.
The $xx$ component $\varepsilon_{\mathrm{i},xx}$ has two prominent peaks at 6.5\,eV and 7.4\,eV while for $\varepsilon_{\mathrm{i},zz}$ the most significant peaks are at 7.2\,eV and 8.4\,eV.
Unfortunately, no experimental data are available for a comparison.

In the case of cubic hafnia, the calculated $\varepsilon_\mathrm{i}$ is very similar to the $\varepsilon_\mathrm{i}$ $xx$ component of t-HfO$_2$, and it is dominated by two sharp excitonic peaks, one at 6.4\,eV, 0.4\,eV below the gap and the second at 7.7\,eV.}
In the case of a-HfO$_2$, where only the RPA calculations were done, a spectral weight shift to higher energies with respect to experimental data can be observed (cf.~\ref{m-eps-compare}), highlighting that the excitonic effects play a major role also in the optical absorption of the amorphous phase.
\changee{
The maximum of $\varepsilon_\mathrm{i}$ for the fast cooled structure is slightly higher, as expected with higher overall mass density and the $\varepsilon_\mathrm{i}$ absorption onset is broader due to the more pronounced effect of the defect states.
No other significant differences between the two generated amorphous structures were predicted.
}}

\section{Conclusions}

\change{
In this work, we have thoroughly tested different parametrisations of the TB-mBJ exchange-correlation potential for predicting optical band gaps and electronic part of the dielectric function of several HfO$_2$ polymorphs.
The crystalline polymorphs, monoclinic, tetragonal and cubic phases, as appearing in the equilibrium phase diagram, were complemented with a supercell model for amorphous material, which is often obtained experimentally due to specific growth conditions.
While the predicted TB-mBJ-orig electronic band gaps were found in excellent agreement with available experimental data, we have shown that such simple comparison is misleading.
Indeed, when comparing the calculated density of electronic states with experimental PES and IPS measurements, the TB-mBJ-semi gives the best results instead.
Regarding the optical properties, we have found that the RPA fares poorly when calculating optical properties of hafnia, as it predicts an overall spectral weight shift to higher energies with respect to experimental data
However, when taking into account the excitonic effects with the BSE theory, we were able to predict dielectric functions resembling closely the measured functions, and we were able to explain the observed spectral features.

In conclusion, the TB-mBJ with the semiconductor parametrisation seems a very viable option for calculating electronic properties and band structure in large HfO$_2$ systems, such as calculations of amorphous cells, defects, surfaces and interfaces.
Regarding the optical properties, the extreme computational costs of the BSE unfortunately mostly eleminates the biggest strength of TB-mBJ, its low computational cost, as with resources sufficient for BSE one might perform a full hybrid or GW calculation instead as well.
}

\ack

This work has been supported by the Ministry of Education, Youth and Sports of the Czech Republic under the project CEITEC 2020 (LQ1601),
by the MOBILITY project 7AMB15AT017 funded by Czech Ministry of Education, Youth and Sports (CMEYS), by the project CZ09/2015 funded by the Austrian Agency for International Cooperation in Education and Research (OeAD-GmbH), and by the Ministry of Education, Youth and Sports from the Large Infrastructures for Research, Experimental Development and Innovations project „IT4Innovations National Supercomputing Center – LM2015070“.
D.H. acknowledges computational resources of the Vienna Scientific Cluster (VSC).

\section*{References}

\bibliographystyle{iopart-num}
\bibliography{bib-db}

\end{document}